# BGP Route Analysis and Management Systems

**Alex A. Stewart** and **Marta F. Antoszkiewicz**
Department of Computer Science
The University of Northern Iowa
305 ITTC Cedar Falls, Iowa 50614-0507
{astewart, mantoszk}@cs.uni.edu

**ABSTRACT**
**The Border Gateway Protocol (BGP) is an important component in today's IP network infrastructure. As the main routing protocol of the Internet, clear understanding of its dynamics is crucial for configuring, diagnosing and debugging Internet routing problems. Despite the increase in the services that BGP provide such as MPLS VPNs, there is no much progress achieved in automating the BGP management tasks.**
**In this paper we discuss some of the problems encountered by network engineers when managing BGP networks. We also describe some of the open source tools and methods that attempt to resolve these issues. Then we present some of the features that, if implemented, will ease BGP management related tasks.**

## I. Introduction

The Internet routing is based on a two-layer routing mechanism where intra-domain routing is achieved through Interior Gateway Protocols (IGPs), and inter-domain routing is done by BGP. BGP is also used as routing protocol for very large, multi-Autonomous System (AS) organizations. At the end, BGP glues all ISPs and their large customers together to form the Internet. Besides traditional routing, BGP also supports routing for new IP services such as Multiprotocol Label Switching Virtual Private Network (MPLS VPN). Unfortunately, while BGP has been designed to meet the high scalability requirements of the Internet and its applications, growth of the Internet and its applications and services made BGP very complex to analyze, manage, and troubleshoot. Some of the reasons that contributed directly to complexity of the BGP manageability are:

- The verboseness of BGP: A single root event such as connectivity change of a BGP router can cause exchanging a huge number of messages between several routers. This number of messages can range from hundreds to minions based on the type of change.
- With the constantly increasing number of routes, the number of paths between these routers and route options increase significantly resulting in rapid growth in BGP routing tables. Managing this huge amount of information is extremely difficult.
- Complexity of configuration options: BGP configuration options can be error-prone and difficult to analyze. There is high possibility that any simple mistake in configuring an attribute such as MEDs or LOCAL-PREFS can cause intricate problems.

Because of the critical scope of BGP's operation, managing BGP and resolving issues caused by BGP misconfiguration can become a black art.

## II. BGP Management Challenges

BGP is primarily used in conjunction with Internet routing, and this is where BGP problems that go unnoticed or that are hard to solve have their heaviest impact. In this section we present some of the BGP-related problems.

### *A. Routing Divergence and Performance Degradation*

Misconfiguring BGP may result in suboptimal routing, which in turn can increase the latency and slow the performance. Furthermore, MED oscillations and flapping routes and can cause routing divergence, increased load on routers and consequently slowing router performance down. Problems related to BGP routing convergence are among the top key issues Internet Service Providers face.

### *A. Security Holes*

BGP is highly vulnerable to wide range of attacks. Such attacks exploit backdoor routes that don't have solid security mechanisms, such as prefix access lists, maliciously and can expose networks to unforeseen major consequences. For example, the attack can result in the backdoor router's peer announcing full Internet routes, which may exhaust the memory of some routers. In addition, dark and murky address space can be used for spam and other attacks. Another example of these security issues are Dark and Murky address spaces.

### *A. Customer-Affecting Disruptions*

While many of the BGP problems affect ISPs, some other issues affect customers. Example of these problems is route peering flaps which may occur on a relatively localized scale and it usually affect few customers. The main consequence of this problem is that it may not be noticed and consequently resolved for long period of time. It is not uncommon for a Tier 1's customer to experience persistent route peering flaps that may occur once a minute or so, for long period of time.

### *B. Customers Impact Service Provider Networks*

In some instances, customer router can have dramatic impact on service provider networks. For example, a misconfiguration of customer router can cause route leakage. Router leakage is injecting the Internet route table back into the Provider Edge (PE) router. However, since PE routers are generally configured with less memory comparing to core routers, a large route leakage can result in memory outage for the PE router, which in turn will cause performance degradation. In some scenarios, route leakages can even cause much worse implications such as inter-domain disruptions. A misconfiguration of tier-1 ISP's customer can cause leakage of thousands of routes that were then announced to the ISP's peers. If one of these peers has a prefix-limit configured and resets the BGP session, severe Internet communication disruption between the two tier-1 ISPs can occur.

### C. *Major Internet Service Disruptions*

Since BGP is the main routing protocol for the Internet, a simple BGP misconfiguration, such as advertising a short AS-PATH artificially to all Internet routes, can cause huge Internet-wide outage. BGP black hole routing routes traffic forwarded to an Internet address under attack to a null address, making the Internet address un-routable. Even though globe-wide Internet routing disruption is uncommon, consequences can be unpredictable and loses can be very huge.

### D. *Unpredictable Bhavior*

BGP interacts with other routing protocols such as IGP that is operating inside ASes. Consequently, the functional complexity of the IP networks increases and can result in unpredictable behavior. This in turn makes root cause analysis and troubleshooting very difficult.

## III. BGP Routing Analysis and Management Tools and Methods

In this we first describe major industry and academic open source BGP management projects aimed at securing the BGP improving its functionality.

### A. *Open source Tools*

- Prefix Sanity Checker (PSC) [14]: PSC is a tool that has been developed by Packet Clearing House; a research institute that conducts research focusing on operations and analysis in the fields of Internet traffic exchange, routing economics, and global network development. The PSC tool aims at helping Internet Service Provider provisioning technicians in validating IP prefixes that customers want to advertise via BGP. It generates correctly formatted prefix-lists that can be copy-pasted into configurations or provisioning systems. The PSC supports lists of IP addresses in several formats and it generates IP prefixes that are compatible with CISCO and Jupiter devices. The produced table of prefixes can be sorted by many attributes such as country or originating AS. The tool flags Prefixes that fail the test and defaults them to not being accepted from the customer.
- Prefix Hijack Alert System (PHAS)[13]: PHAS is a light-weight, easy to implement and deploy, notification system that has been developed by Colorado State University. PHAS provides several two main functionalities: (1) alerting prefix owners when their BGP origin changes. PHAS is real-time system and it provides timely and reliable notification when origin AS changes. Thus, allowing prefix owners to quickly detect and respond to prefix hijacking attempts; (2) protecting against false BGP origins. PHAS provides the functionality of detecting prefix hijacking events that involve announcing more specific prefixes or modifying the last hop in the path.
- Route Views project [15]: This project has been developed by the University's University of Oregon. The main goal of this system is to help ISP operators obtaining real-time information about major

Internet backbones. One of the major restrictions of Route View is that, unlike previously described tools, it either provides a partial view of the routing system or it doesn't provide real-time access to routing data.

### B. *BGP Route Analysis in Literature*

One of the first works to address BGP route analysis and convergence problems was [16]. The authors showed that EBGP diverging problem can be caused by routing policies. Griffin and Wilfong [6] used graph theory to perform detailed analysis of EBGP convergence causes and properties. They namely proved that deciding whether a given EBGP configuration can converge is an NP-Complete problem.

Followed by NP-completeness proof of EBGP convergence, several solutions have been proposed to resolve the EBGP convergence problem. Govindan et al. [4] described a static solution based on analyzing routing policies statically by software tools. These tools then can determine whether EBGP policy conflicts could lead to a divergence. Villamizar et al. [17] proposed a dynamic mechanism to prevent route update storms. Their solution uses control the dissemination of routing updates based on route flap dampening.

Several problems with route reflection in IBGP have been also investigated. Dube and Scudder [2] showed that route reflection misconfiguration can cause routing loops or incorrect routing decisions. They also proposed solutions to avoid such problems. The authors in [3] proved that discovering whether an IBGP route reflector configuration will cause forwarding loops is an NP-complete problem and provided method for safe route reflector configuration. Routing oscillation is another IBGP routing problem that has been studied extensively. Routing oscillation occurs in networks running IBGP and configured with route reflection or confederations. The routing oscillation was first discussed by Cisco systems [1]. The problem was further investigated in [11] and the authors proposed a modifying the IBGP to handle this problem [18]. However, the solution proposed in [18] failed to eliminate persistent oscillations. Labovitz et al. [9] studied the adverse effects of IBGP route oscillations and used real routing traffic traces to describe a wide range of anomalous behavior in IBGP routing. Other researchers focused on analyzing the cause of these instabilities and suggested solutions [10].

### C. *Next Generation BGP Management System*

An effective BGP management system should have BGP routing analysis and troubleshooting integrated features. In this subsection we describe some services that should be provided by a successful next generation BGP management system to be able to deliver solid BGP management tasks for enterprise networks and service providers.

#### 1) *Framework for BGP Routing Table Visualization*

Ability to visualize BGP routing tables can provide network operators with a static view of the current BGP topology. Such complete view of the network exhibiting number of prefixes that are carried by routing branches as well as numerical counts will allow network engineers to see the behavior of the BGP network at one glance. This will obviously make it easier to evaluate the current configurations and assess existing routing policies. Filtering mechanism should be provided so that network engineers can select how detailed the network topology will be.

### *2) BGP Routing Historical Analysis*

Assembling precise historical view of BGP routing activity helps significantly in troubleshooting BGP routing issues. There are many route analytics in market that provide different levels of historical views. However, providing accurate historical view is one of biggest challenges since such view should provide detailed accurate view of all BGP activities at any moment. Furthermore, such tools should help in analyzing the impact of each of these changes on the ISP and the global Internet routing. Figure 1 depicts a sample screenshot of a high-level BGP routing history analysis.

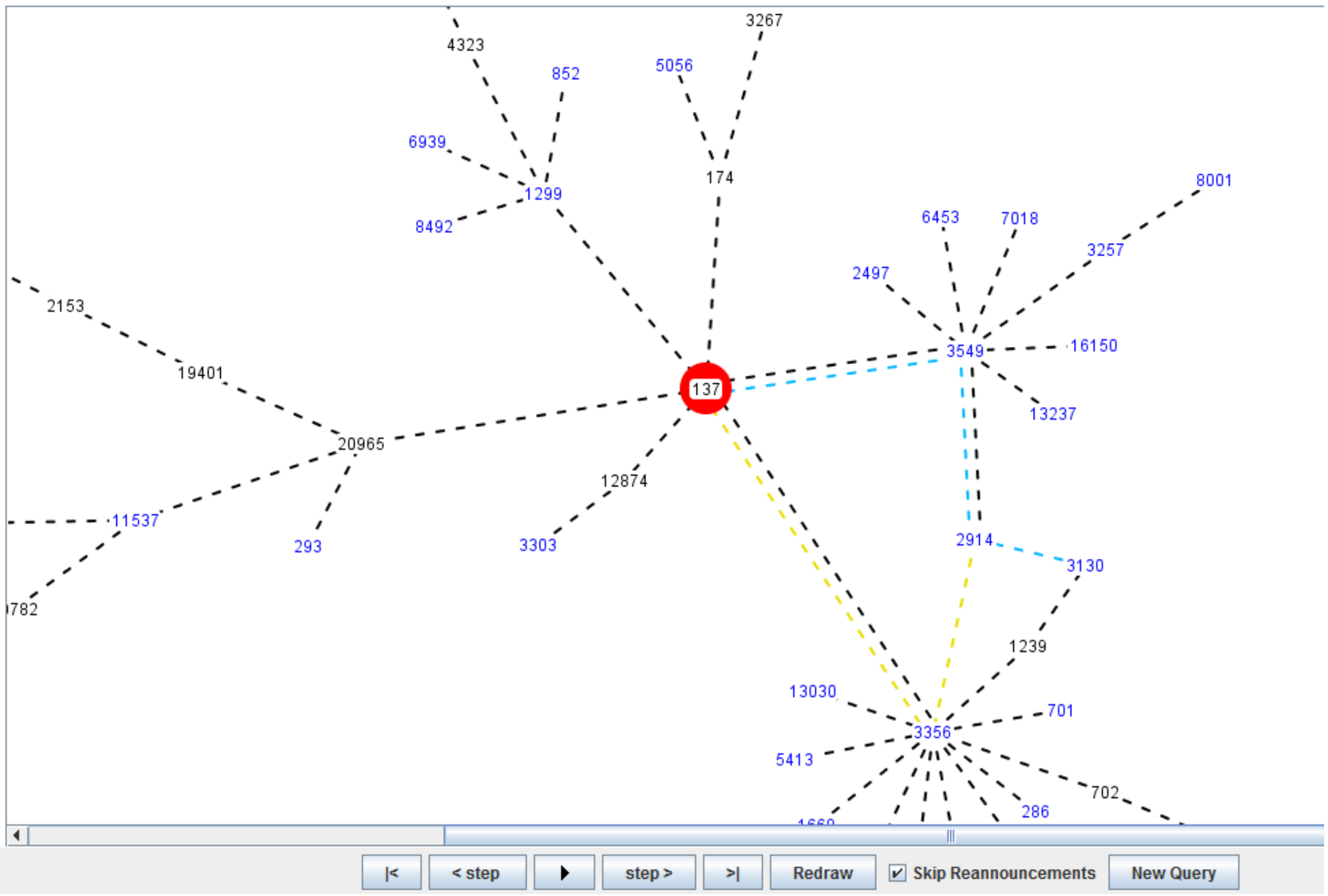

**Figure 1:** BGP routing history analysis can provide engineers with real-time detailed information about changes in routes.

### *3) BGP Root Cause Analysis*

The Border Gateway Protocol (BGP) is a verbose protocol, which makes it extremely difficult for network operators and engineers to perform root cause analysis of its dynamics. Because of the lack to clear understanding of the dynamics of the BGP, network engineers' ability to address BGP's shortcomings is largely restricted. A successful BGP management system should have the capability to determine the cause of any routing change, and where did the change originate.

## V. Conclusion

BGP plays critical role in Internet routing and there is a significant need for a tool to automate its tedious manual configuration and troubleshooting techniques. In this paper we described some of the open source tools made available by academia in attempt to automate managing BGP management tasks. We also described some of the algorithms proposed to resolve some of the major issues facing BGP safe configuration. Finally, we described some of the features that should be provided in future BGP management tools to help network engineers in preventing, understanding and quickly solving BGP routing problems.